\documentclass[lettersize,journal]{IEEEtran}
\usepackage{amsmath,amsfonts}
\usepackage{algorithmic}
\usepackage{array}
\usepackage[caption=false,font=normalsize,labelfont=sf,textfont=sf]{subfig}
\usepackage{textcomp}
\usepackage{stfloats}
\usepackage{url}
\usepackage{cite}
\usepackage{verbatim}
\usepackage{graphicx}
\usepackage[colorlinks=true,linkcolor=blue,urlcolor=blue,citecolor=blue]{hyperref}

\usepackage[colorinlistoftodos]{todonotes}

\def\BibTeX{{\rm B\kern-.05em{\sc i\kern-.025em b}\kern-.08em
    T\kern-.1667em\lower.7ex\hbox{E}\kern-.125emX}}
\usepackage{balance}
\usepackage[export]{adjustbox}
\begin{document}
\title{\bf Self-Error Correcting Method for Magnetic-Array-Type Current Sensors
in Multi-Core Cable Applications}
\author{Xiaohu Liu, Keyu Hou, Kang Ma, Jian Liu, Angang Zheng, Zhengwei Qu, Wei Zhao, Lisha Peng, {\it Member, IEEE},  Songling Huang, {\it Senior Member,  IEEE}, Shisong Li$^\dagger$, {\it Senior Member, IEEE} 
\thanks{Xiaohu Liu, Kang Ma, Jian Liu, Lisha Peng, Songling Huang and Shisong Li are with the Department of Electrical Engineering, Tsinghua University, Beijing 100084, China. Wei Zhao is with the Department of Electrical Engineering, Tsinghua University, Beijing 100084, China, and also with the Yangtze Delta Region Institute of Tsinghua University, Jiaxing, Zhejiang 314006, China. Keyu Hou and Zhengwei Qu are with the School of Electrical Engineering, Yanshan University, Qinhuangdao 066004, China. Angang Zheng is with the China Electric Power Research Institute, Beijing 100192, China.}
\thanks{This work was supported by the Smart Grid-National Science and Technology Major Project under Grant No. 2024ZD0803305 and National Natural Science Foundation of China under Grant No. 52507014.}
\thanks{$\dagger$Email: shisongli@tsinghua.edu.cn}}

\markboth{}{}

\maketitle

\begin{abstract}
Data-driven methods enable online assessment of error states in magnetic-array-type current sensors, and long-term measurement stability can be enhanced through further self-error correction. However, when the magnetic-array-type current sensors are applied to multi-conductor systems such as multi-core cables, the time-varying correlations among conductor currents may degrade the performance of multi-latent-variable data-driven models for error evaluation. To address this issue, this paper proposes a robust self-error correcting method for magnetic-array-type current sensors even under significant variations in phase current correlations (e.g., large fluctuations in three-phase current imbalance). By incorporating phase current decoupling and principal component analysis (PCA), the correlation analysis of multi-latent variables (i.e., multi-conductor currents) is transformed into a single-latent-variable (corresponding to single phase current) modeling problem. Experimental results demonstrate that the proposed method effectively detects error drifts of magnetic field sensors as low as $2\times10^{-3}$ in relative error and $2\times10^{-3}$\,rad in phase error. Accurate evaluation and correction of each magnetic field sensor’s error drifts substantially eliminates the overall error drift in the magnetic-array-type current sensor, validating the feasibility and effectiveness of the proposed self-error correcting method.
\end{abstract}


\section{Introduction}

\IEEEPARstart{M}{ulti-core cables} are pivotal in medium- and low-voltage urban power grids and industrial distribution systems, boosting power transmission/distribution reliability with advantages of space efficiency and easy installation~\cite{Liu2025Research1,Liu2023Online2}. Accurate measurement of individual conductor current in multi-core cables is critical for load balancing, overload prevention, and power distribution optimization. However, conventional techniques (e.g., electromagnetic current transformers, Rogowski coils, zero-flux current sensors)~\cite{Ripka2010Electric3,Sun2024High4} require stripping the cable sheath for separate conductor current measurement, posing significant barriers to distributed and wide-area current sensing for multi-core cables.

Magnetic-array-type current sensors can achieve non-contact current measurement for multi-core cables without cable sheath stripping. This type of current sensors detects the cable surface magnetic fields and directly computes the individual conductor current through inverse calculations~\cite{Zhu2017onsite,Luo2024Research13,Liu2022Coreless7,Liu2021Nonintrusive10}, which have been extended to current measurement scenarios like overhead transmission lines~\cite{Chen2024Contactless,Chen2022Intelligent,Wu2019Overhead}, rectangular busbars ~\cite{Ma2025Eliptical,George2023Rectangular,Li2021Wideband}, and even multi-strand superconductor cables~\cite{Bellina2002Optimization}. 

Measurement stability serves as the cardinal performance metric for quantifying an instrument's long-term operational reliability. To accurately reconstruct the phase current in multi-core cables, it is essential to ensure the measurement accuracy of all magnetic field sensors. However, environmental factors (e.g., temperature) can cause significant error drifts in some magnetic field sensors, thereby degrading current measurement accuracy. The long-term stability of current measurement can be significantly improved if the error-prone magnetic field sensors can be self-corrected online. 

To evaluate the error state of the magnetic-array-type current sensor online, a calibration method based on physical standards can be employed. For instance, a standard current sensor capable of live-line installation may be used to calibrate the magnetic-array-type current sensor. However, the implementation requires the design of an online installation standard current sensor~\cite{Li2016High}, making the hardware structure more complicated and costly. An alternative solution is to deploy standard magnetic field-generating coils at each magnetic field sensor location to excite a standard magnetic field at a frequency different from that of the power line. However, this method requires precise coil size design and a highly stable reference current source, thereby increasing the complexity and cost of the current sensing system.

The data-driven approach shows significant promise for the online evaluation of error states in magnetic-array-type current sensors. Through a self-supervised strategy, it can perform self-error correction for each magnetic field sensor, providing the notable advantages of high sensitivity and low cost. The efficacy of those methods has been demonstrated in various scenarios, including industrial process monitoring~\cite{Chen2025New, Yu2025Challenges} and metrology performance evaluation for energy metering instruments~\cite{Tong2024novel, Zhang2023physics}.

For industrial process monitoring, in~\cite{Kong2022deep}, a deep PCA-ICA latent variable model is presented and its effectiveness is validated through the Tennessee Eastman process. A. Gandhimathinathan et al.~\cite{Gand2024Sensor} propose a deep learning approach and achieved a 98.2\% recognition rate for anomalous error states of sensors in a nuclear power plant.

For metrology performance evaluation of electricity metering equipment, in~\cite{Xia2023Measurement}, an improved backpropagation neural network-based parameter estimation method is adopted to solve measurement error parameters of distributed smart meters. In~\cite{Zhang2017Monitoring}, under the condition that the three-phase voltage imbalance remains relatively stable, a PCA-based method successfully achieves error status evaluation of 0.2-class CVTs. However, in practical operation, the correlation constraints of the three-phase voltage are often violated due to supply fluctuations and time-varying loads, ultimately compromising the accuracy and sensitivity of the proposed data-driven method.

The essence of data-driven methods is to ensure that the trained model consistently matches the correlation constraints among latent variables throughout the monitoring process. In~\cite{Lu2018Data}, an adaptive multi-mode method is proposed to achieve adaptive matching across multiple modes. Due to the stochastic and time-varying nature of three-phase loads, the correlation among phase currents in multi-core cables may vary continuously, making it infeasible to accurately model these correlations using a limited set of discrete modes. In~\cite{Zhang2019Decting}, a rigid constraint condition is established based on the measurement consistency of co-phase CVTs, and then a PCA-based single-latent-variable model is constructed, which significantly enhances the error state evaluation accuracy for CVTs. 

The feasibility of applying data-driven methods to the online error evaluation of magnetic-array-type current sensors is first explored in~\cite{Liu2024self} for single-phase current measurement. A single-latent-variable model is developed based on PCA, in which measurements of all magnetic field sensors are correlated solely with the single-phase current. This model enables self-error correction through online detection and elimination of error-drifted magnetic field sensors. When magnetic-array-type current sensors are applied to multi-core cable current measurement, the dynamic correlation variations among phase currents pose challenges to error evaluation. 

Inspired by \cite{Lu2018Data,Zhang2019Decting}, if the multi-latent-variable correlation analysis can be transformed into a single-latent-variable modeling framework, the resulting error evaluation model could avoid interference from time-varying multi-latent-variable correlations. Based on the above, the innovations of this paper are summarized as follows:

\begin{itemize}
    \item This paper proposes to transform the correlation analysis of measurements by the magnetic sensor array into a correlation analysis of in-phase currents derived from different sub-combinations of magnetic field sensors. Even under significant variations in phase current correlations, the proposed method can robustly detect error-drifted magnetic field sensors and quantify the error drift magnitude.
    
    \item An identification method for error-drifted magnetic field sensors is proposed. Each error-drifted magnetic field sensor is determined by iteratively removing it and assessing whether the variation patterns of the $Q$ statistic curves for in-phase current amplitudes and phases change significantly. 

    \item The evaluation of the drifts in the relative and phase errors of magnetic field sensors is transformed into a bi-objective optimization problem. By incorporating intelligent evolutionary algorithms, the error drift amounts of all magnetic field sensors are solved, thereby achieving self-error correction of magnetic-array-type current sensors.
\end{itemize}

The remainder of this article is organized as follows.
Section \ref{sec:02} introduces the proposed method for self-error correction of magnetic-array-type current sensors. Section \ref{sec:03} presents the experimental study aimed at demonstrating the validation of the proposed data-driven approach. Finally, the conclusion is given in Section \ref{sec:04}.

\section{Method for Self-Error Correcting Under Dynamic Variations in Multi-Conductor Current Correlation}
\label{sec:02}

\subsection{Self-Error Correcting Principle of Magnetic-Array-Type Current Sensors Applied to Multi-Core Cable Current Measurement}\label{sec:II-A}

\begin{figure}[tp!]
\centering
\includegraphics[width=0.7\columnwidth]{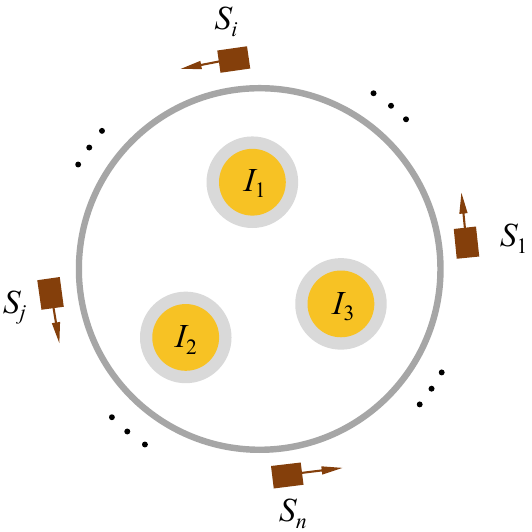}
\caption{Non-contact current measurement for multi-core cables using a magnetic sensor array, with rectangles representing magnetic field sensors and arrows indicating their sensitivity orientations.} 
\label{fig:magnetic-array}
\end{figure}

\begin{figure}[!t]
\centering
\includegraphics[width=\columnwidth]{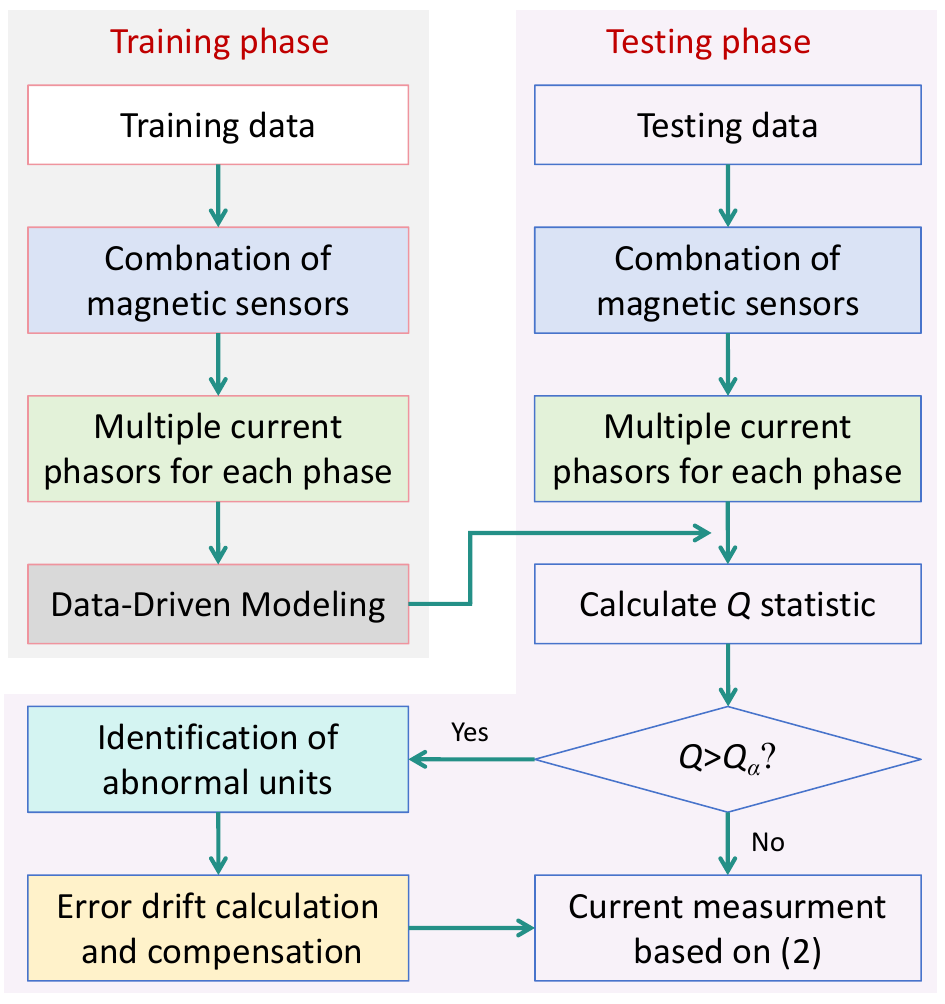} 
\caption{Principle of the proposed self-error correcting method for magnetic-array-type current sensors applied to multi-core cables.} 
\label{fig:Diagram-flow}
\end{figure}

As shown in Fig.~\ref{fig:magnetic-array}, as an example, a magnetic sensor array is used for three-core three-phase current measurement. In the long-term current measurement, the correlation between phase currents could vary significantly due to the random phase load changing, resulting in the linear correlation variation between the measurements of all magnetic field sensors. Directly applying a traditional multi-latent-variable model to analyze the magnetic sensor array measurements proves inadequate for achieving satisfactory performance. To achieve self-error correcting of magnetic field sensors under significant time-variation linear correlation for multi-conductor currents, this paper presents a data-driven self-error correcting method that analyzes linear correlations in derived multiple currents per phase using measurements from different magnetic field sensor sub-combinations. As shown in Fig.~\ref{fig:Diagram-flow}, the proposed method consists primarily of a training and testing phase.

For the training phase, all magnetic field sensors operate under normal error conditions, and their measurements are collected as training data. These magnetic field sensors are grouped into multiple sub-combinations, and the three-phase fundamental current phasors can be derived from each sub-combination's measurements. PCA is then applied to construct a data-driven model for the amplitude and phase of these current phasors per phase.

In the testing phase, all magnetic field sensors are similarly grouped to obtain multiple fundamental current phasors for each phase. Based on the constructed data-driven model in the training phase, the $Q$ statistic for both amplitude and phase of the derived in-phase current phasors is calculated. If the $Q$ statistic exceeds the predefined threshold, it indicates the presence of error-drifted magnetic field sensors in the magnetic sensor array. All the error-drifted magnetic field sensors are identified, and their error drift values are then quantified for error compensation through an error estimation algorithm. The phase currents are subsequently recalculated using the compensated magnetic field measurements to ensure current measurement accuracy. In cases where the $Q$ statistic remains within the acceptable range, the phase currents are computed directly from the raw magnetic field sensor measurements without specific error correction. The detailed implementation methodology will be systematically presented in subsequent sections.

\subsection{Phase Current Linear Correlation Variation Analysis Caused by Error-Drifted Magnetic Field Sensors}\label{sec:II-B}

As illustrated in Fig.~\ref{fig:magnetic-array}, the three-phase current phasors and the output voltage phasor of each magnetic field sensor exhibit the following relationship 

\begin{equation}\label{eq:1}
\begin{bmatrix}
\dot{V}_1 \\
\vdots \\
\dot{V}_n \\
\vdots \\
\dot{V}_{N}
\end{bmatrix}
=
\begin{bmatrix}
k_{11} & k_{12} & k_{13} \\
\vdots & \vdots & \vdots \\
k_{n1} & k_{n2} & k_{n3} \\
\vdots & \vdots & \vdots \\
k_{N1} & k_{N2} & k_{N3}
\end{bmatrix}
\begin{bmatrix}
\dot{I}_1 \\
\dot{I}_2 \\
\dot{I}_3
\end{bmatrix}
\end{equation}
where \(\dot{I}_1\), \(\dot{I}_2\), \(\dot{I}_3\) respectively represents fundamental current phasor of phase a, b, and c. $\dot{V}_n$ denotes the fundamental voltage phasor of the $n$th magnetic field sensor, $N$ is the total number of magnetic field sensors. The coefficient $k_{nj}$ quantifies the complex proportionality between the output voltage phasor $\dot{V}_n$ and the current phasor $\dot{I}_j$ ($j$=1, 2, 3) when only the $j$th conductor carries current. The complete coefficient matrix formed by all $k_{nj}$ elements is defined as the transfer matrix $\mathbf{K}$.

Following the calibration and determination of the transfer matrix $\mathbf{K}$, the current phasor for each phase are reconstructed based on the measured voltage phasors from all magnetic field sensors, according to the governing equation as follows
\begin{equation}\label{eq:current-inversion}
    \boldsymbol{I}=\left(\mathbf{K}^{\mathrm{T}} \mathbf{K}\right)^{-1} \mathbf{K}^{\mathrm{T}} \boldsymbol{V} 
\end{equation}
where $\boldsymbol{I}$ and $\boldsymbol{V}$ are column vectors composed of the current phasors and the output voltage phasors of the magnetic field sensors, respectively. Furthermore, we define the magnetic field-current correlation matrix as $\mathbf{M}=\left(\mathbf{K}^{\mathrm{T}} \mathbf{K}\right)^{-1} \mathbf{K}^{\mathrm{T}}$.

Using the measurements from any three magnetic field sensors in the magnetic-array-type current sensor, the phase current phasors can be calculated according to~(\ref{eq:current-inversion}). Taking the $g$-th, $h$-th, $l$-th, and $s$-th magnetic field sensors, for instance, they are grouped into sub-combinations of three, and the phase currents can be further derived using the measurements of each sub-combination. If the $g$-th magnetic field sensor exhibits an error drift of $\Delta\epsilon_g$, the resulting error drift in phase-a current for each sub-combination can be expressed as
\begin{equation}\label{eq:current-error-drift}
\begin{cases}
\begin{aligned}
\Delta \dot{I}_{1}^{g,h,s} &= m_{11}^{g,h,s} \dot{V_{1}} \Delta \varepsilon_{g} \\
\Delta \dot{I}_{1}^{g,h,l} &= m_{11}^{g,h,l} \dot{V_{1}} \Delta \varepsilon_{g} \\
\Delta \dot{I}_{1}^{g,l,s} &= m_{11}^{g,l,s} \dot{V_{1}} \Delta \varepsilon_{g} \\
\Delta \dot{I}_{1}^{h,l,s} &= 0
\end{aligned}
\end{cases}
\end{equation}
where $\Delta \dot{I}_{1}^{g,h,s}$ represents the error drift of the phase-a current derived from the measurements of the $g$-th, $h$-th, and $s$-th magnetic field sensors. $m_{11}^{g,h,s}$ denotes the element in the first row and first column of the magnetic field-current coefficient matrix corresponding to the magnetic sensor array formed by the $g$-th, $h$-th, and $s$-th magnetic field sensors. Other symbols follow similar definitions.

The magnetic field-current coefficient matrix varies across different sub-combinations due to the different spatial arrangements of magnetic field sensors. This implies that the coefficients $m_{11}^{g,h,s}$, $m_{11}^{g,h,l}$, $m_{11}^{g,l,s}$ in~(\ref{eq:current-error-drift}) are unequal, leading to different error drifts in the computed phase-a currents, indicating the linear correlation variation between the derived in-phase currents. Thus, specific error-drifted magnetic field sensors can be identified by analyzing variations in the linear correlation of in-phase currents obtained from different sub-combinations of magnetic field sensors.

\subsection{Evaluation of Measurement Error Status of Magnetic-Array-Type Current Sensor by Detecting Phase Current Linear Correlation Variation}\label{sec:II-C}

Principal component analysis is an effective method for handling linear correlation analysis problems, capable of transforming linearly correlated high-dimensional variables into a few independent principal components that characterize the data features~\cite{greenacre2022principal}. Based on this method, the measurement data composed of several data channels operating in a non-error-drifted state are used as training data to learn the principal component subspace $\mathbf{P}$, with specific implementation details provided in reference~\cite{Liu2024self}. The data-driven modeling of the column vector $\boldsymbol{x}$ composed of measurement data from different data channels is given by the following equation
\begin{equation}\label{eq:data-driven-model}
    \tilde{\boldsymbol{x}} = \boldsymbol{x}\mathbf{PP}^\text{T}
\end{equation}
where $\tilde{\boldsymbol{x}}$ represents the projection of $\boldsymbol{x}$ onto the principal component subspace.

When none of the data channels exhibit an error drift, the linear correlations among the elements in $\boldsymbol{x}$ remain consistent, and the $\tilde{\boldsymbol{x}}$ computed from (\ref{eq:data-driven-model}) will be approximately equal to $\boldsymbol{x}$. Otherwise, significant deviations between $\boldsymbol{x}$ and $\tilde{\boldsymbol{x}}$ will occur. The $Q$ statistic, also known as the Squared Prediction Error, quantifies the squared length of the projection of an observed data vector onto the residual subspace. Essentially, it serves as a quantitative measure of the discrepancy between the observed data and the predicted values by the data-driven model. The $Q$-statistic is employed to quantify such variations in linear correlation, defined as
\begin{equation}\label{eq:Q-calculation}
    Q = \|\tilde{\boldsymbol{x}} - \boldsymbol{x}\|_{2}^{2}
\end{equation}

When the $Q$ statistic significantly exceeds its control threshold $Q_{\text{ctr}}$, it indicates a change in the linear correlations. The definition of $Q_{\text{ctr}}$ is  as follow~\cite{Jackson1979control}
\begin{equation}\label{eq:5}
Q_{\text{ctr}}^2 = \theta_1 \left[ \frac{C_\beta \sqrt{2\theta_2 h_0^2}}{\theta_1} + 1 + \frac{\theta_2 h_0 (h_0 - 1)}{\theta_1} \right]^{1/h_0}  
\end{equation}
where $\theta_i=\sum_{j=m+1}^{N} (\sigma_j)^i (i=1,2,3)$, $h_0=1-2\theta_1\theta_3/(3\theta_2^2)$, $C_\beta$ is the critical value of the normal distribution at a significance level of {\color{blue}$\beta=0.99$}, and $m$ denotes the number of principal elements, which can be obtained using cumulative variance contribution rate method.

\begin{figure}[!t]
\centering
\includegraphics[width=\columnwidth]{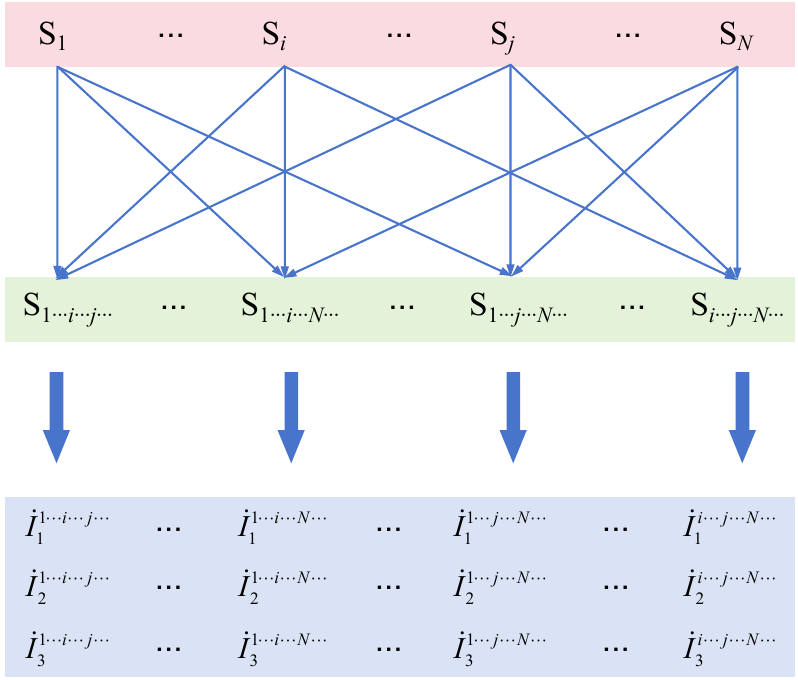} 
\caption{Illustration of phase currents derivation from measurements of different sub-combinations of magnetic field sensors.} 
\label{fig:current-phasors-dec}
\end{figure}

The analysis in Section~\ref{sec:II-B} indicates that error drifts of magnetic field sensors may disrupt the linear correlation among multiple current values of a certain phase, which are calculated from measurement data of different magnetic field sensor sub-combinations. Accordingly, this paper proposes to achieve an error status evaluation of magnetic-array-type current sensor by detecting the linear correlation changes between the multiple currents for each phase. As shown in Fig.~\ref{fig:current-phasors-dec}, all magnetic field sensors are grouped into  $C_N^n$ sub-combinations, where $C_N^n$ represents the number of possible $n$-element ($3 \leq n < N$) sub-combinations that are selected from a set of $N$ distinct magnetic field sensors. $\textbf{S}_{1\cdots i \cdots j \cdots}$ denotes a sub-combination containing $n$ magnetic field sensors, including the $1{\text{st}}$, $i{\text{th}}$, and $j{\text{th}}$ magnetic field sensor. Using the output voltage phasors from all sub-combinations, $C_N^n$ currents per phase are obtained, such as the phase-$a$ currents $\dot{I}_{1}^{1\cdots i \cdots j \cdots}, \ldots, \dot{I}_{1}^{i\cdots j \cdots N \cdots}$.

If a linear correlation change is detected among the derived currents for a certain phase, the $Q$ statistic for the amplitude or phase of these currents is checked against its control threshold. This allows for evaluating whether any magnetic field sensor exhibits an error drift in the magnetic-array-type current sensor.

\subsection{Identification of Error-Drifted Magnetic Field Sensors by Comparing $Q$ Statistic Curves Variation Trend}\label{sec:II-D}

\begin{figure}[!t]
\centering
\includegraphics[width=\columnwidth]{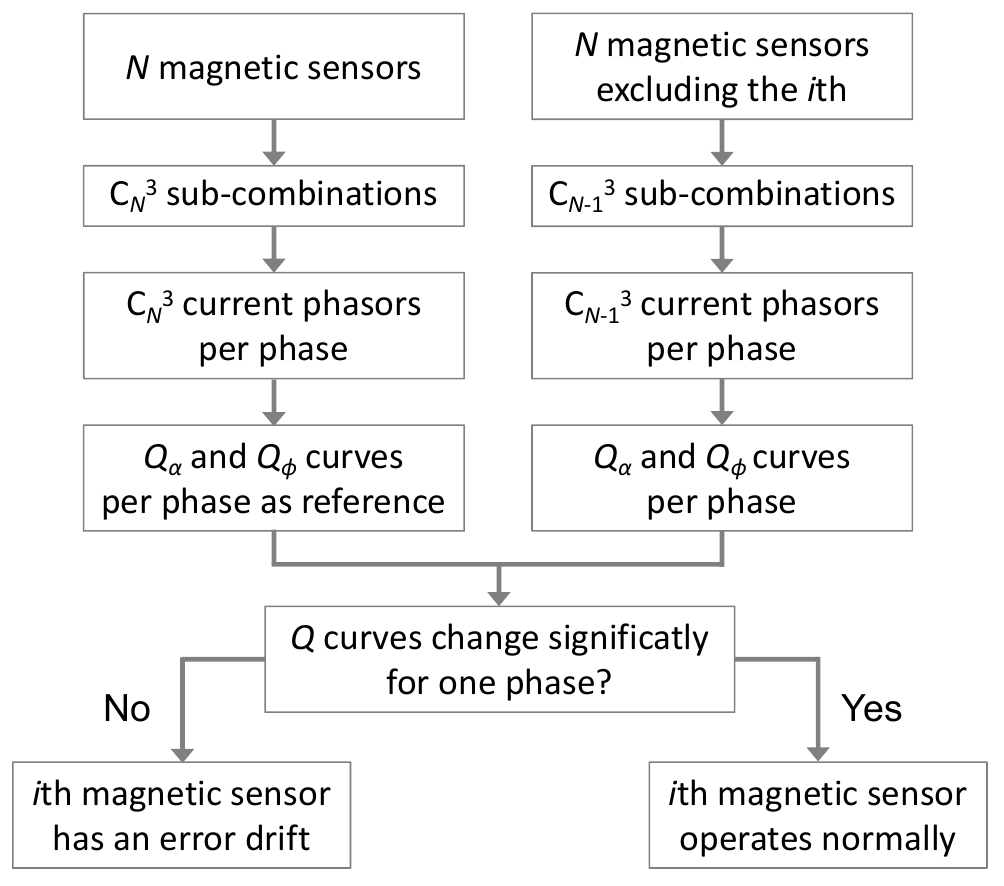} 
\caption{Flow chart for identification of the error-drifted magnetic field sensors.} 
\label{fig:flowchart_error_drifted_identification}
\end{figure}

Upon detecting an abnormal error state of the magnetic-array-type current sensor, further discrimination of error-drifted magnetic field sensors is required. Given the heterogeneous error drift characteristics of magnetic field sensors and the stochastic, independent nature of these drifts, magnetic field sensors with abnormal error states typically exhibit distinct drift patterns and magnitudes. The variation trends of the $Q$ statistic essentially reflect the error drift patterns of magnetic field sensors. Any error drift occurring in a magnetic field sensor will inevitably manifest as significant alterations in the $Q$ statistic variation trend. In this study, we employ a sequential exclusion method, systematically eliminating individual magnetic field sensors to assess their impact on the morphological characteristics of the $Q$ statistic curve, thereby enabling the identification of error status in the individually excluded magnetic field sensor. The identification of error-drifted magnetic field sensors includes the following steps, which is presented in Fig.~\ref{fig:flowchart_error_drifted_identification}.

First, all magnetic field sensors are grouped into combinations of three, yielding $C_N^3$ sub-combinations. For each phase, a total number of $C_N^3$ fundamental current phasors can be obtained using the output fundamental voltage phasors of the corresponding sub-combinations according to (\ref{eq:current-inversion}). Subsequently, the $Q$ statistic curves of either the amplitude ($Q_\alpha$) or phase ($Q_\phi$) of these $C_N^3$ fundamental current phasors are computed for each phase using (\ref{eq:Q-calculation}). These resulting $Q$ statistic curves are then employed as reference.

Subsequently, each magnetic field sensor is sequentially excluded from the magnetic sensor array, and the remaining $N-1$ magnetic field sensors are grouped into sub-combinations of three, yielding $C_{N-1}^3$ sub-combinations. For each phase, the fundamental current phasor corresponding to each of the $C_{N-1}^3$ sub-combinations is calculated according to (\ref{eq:current-inversion}). Further, the $Q$ statistic curves of the fundamental current amplitude and phase for each phase are computed based on (\ref{eq:Q-calculation}). Comparing the variation trend of these curves with the reference curves obtained in the first step, if the variation trend remains essentially unchanged, the excluded magnetic field sensor is identified to be in a normal error state, otherwise, it is determined to exhibit an error drift.

\subsection{Quantification of the Error Drifts of Magnetic Field Sensors by Solving a Bi-Objective Optimization Problem}\label{sec:II-E}

After localizing all magnetic field sensors with an abnormal error state, a $Q$ statistic minimization-based approach is proposed for quantification of the error drift.

For any magnetic field sensor in an abnormal error state, let $\Delta\alpha$ and $\Delta\phi$ represent the unknown magnitude and phase error drifts, respectively. Each of the identified error-drifted magnetic field sensors and all magnetic field sensors in a normal error status are grouped into sub-combinations of three. Using (\ref{eq:current-inversion}), the magnetic field measurements from each sub-combination are used to compute the phase currents. When a sub-combination contains the error-drifted magnetic field sensor, its measurement data is first compensated using $\Delta\alpha$ and $\Delta\phi$ before performing phase current calculation. This process yields multiple fundamental current phasors for each phase, with the number of current phasors being equal to that of the sub-combinations.

The $Q$ statistics of both amplitude and phase are calculated for the multiple current phasors of each phase, denoted as $Q_{\alpha,j}$ and $Q_{\phi,j}$, respectively. These phase-specific $Q$ statistics are then summed across all phases, i.e.,

\begin{equation}\label{eq:bi-objective}
\begin{cases} 
Q_\alpha(\Delta\alpha, \Delta\phi) = \sum_{j=1,2,3} Q_{\alpha, j}(\Delta\alpha, \Delta\phi) \\ 
Q_\phi(\Delta\alpha, \Delta\phi) = \sum_{j=1,2,3} Q_{\phi, j}(\Delta\alpha, \Delta\phi)
\end{cases}
\end{equation}

From (\ref{eq:bi-objective}), the values of $Q_\alpha$ and $Q_\phi$ are determined by the error drifts of both amplitude and phase of the abnormal magnetic field sensor. When $\Delta\alpha$ and $\Delta\phi$ can effectively compensate for the magnetic field sensor error drifts, the linear correlation among multiple current phasors for each phase will be minimized. Therefore, the objective of evaluating the error drifts is to minimize $Q_\alpha$ and $Q_\phi$, which leads to the following bi-objective optimization problem

\begin{equation}\label{eq:bi-objective-problem}
\left\{
\begin{array}{l}
\text{Minimize } \left[Q_{\alpha}(\Delta \alpha, \Delta \phi), Q_{\phi}(\Delta \alpha, \Delta \phi)\right] \\
\text{s.t. } \quad |\Delta \alpha| \leq \alpha_{s}, \\
\text{s.t. } \quad |\Delta \phi| \leq \phi_{s}
\end{array}\right.
\end{equation}

Multi-objective optimization methods are primarily categorized into traditional approaches and modern intelligent optimization techniques. The fundamental principle of traditional methods involves transforming multi-objective functions into single-objective functions, thereby enabling efficient solutions through techniques such as the gradient descent method. However, these methods are more suited to scenarios where the objective functions are well-defined analytically and require careful weight selection. Modern intelligent optimization methods represent the mainstream approach to multi-objective optimization, being suitable for optimization problems where the objective function is non-analytic. The objective function in this paper, following phase current decoupling and principal component decomposition steps, is non-analytic.

Thus, the non-dominated sorting genetic algorithm (NSGA)-based multi-objective optimization method is used to solve the problem formulated in (\ref{eq:bi-objective-problem}). The algorithm effectively handles conflicts between objective functions through non-dominated sorting and crowding distance computation, yielding a well-distributed Pareto optimal solution set without requiring gradient information. Upon obtaining the error drift parameters $(\Delta\alpha,\Delta\phi)$ for the magnetic field sensor with abnormal error status and applying compensation, the measured current is reconstructed according to (\ref{eq:current-inversion}), thereby achieving self-error correction of the magnetic-array-type current sensor.

Analysis indicates that, when applying the proposed method for self-error correction of magnetic-array-type current sensors in a three-phase current measurement system, the magnetic array must contain at least five magnetic field sensors, with at least three magnetic field sensors working in their normal status. To achieve better self-error correction performance, the magnetic field sensors in a normal error status should preferably be uniformly distributed around the multi-core cable, and this point will be demonstrated in Section \ref{sec:III-E}.

\section{Experimental Verification}
\label{sec:03}
\subsection{Experimental Test Platform}\label{sec:III-A}

\begin{figure}[!t]
\centering
\includegraphics[width=1\columnwidth, height=0.5\textheight, keepaspectratio]{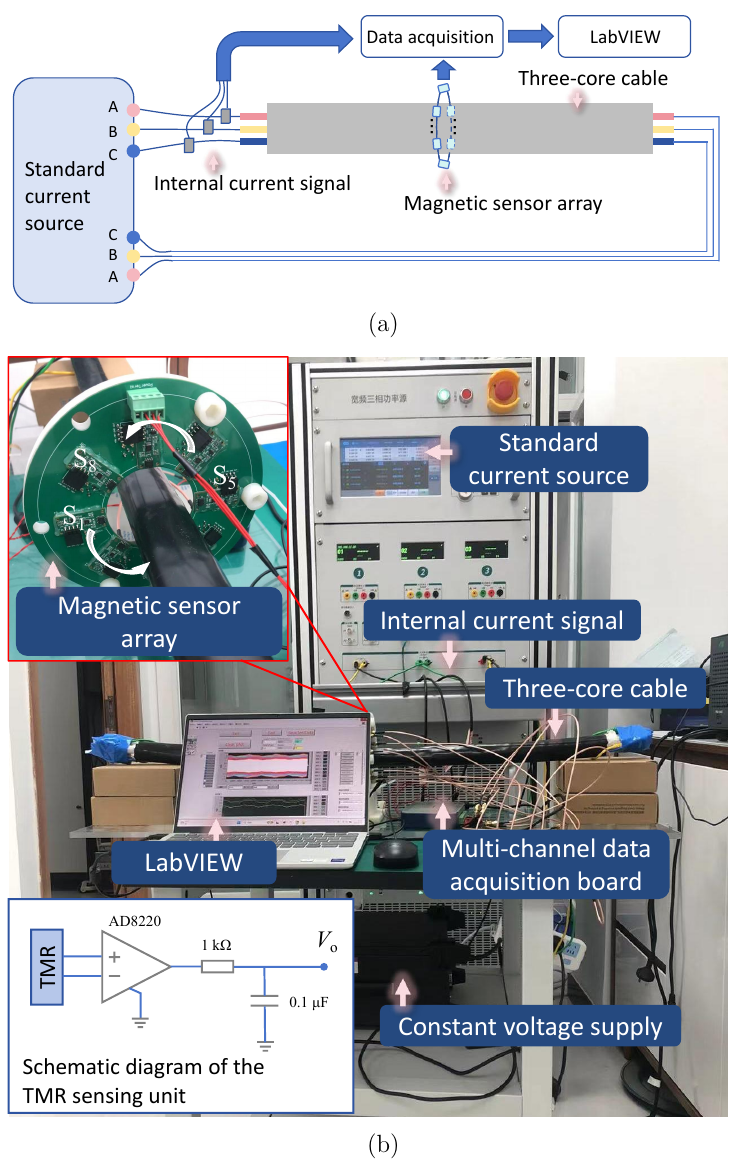} 
\caption{Schematic of the self-error correction experiment and corresponding physical test platform (b).} 
\label{fig:experiment-platform}
\end{figure}

To validate the effectiveness of the proposed self-error correcting method, experimental verification is performed on a magnetic-array-type current sensor applied for current measurement of a three-phase three-core cable. The overall schematic of the current measurement experiment and corresponding physical test platform are illustrated in Fig.~\ref{fig:experiment-platform}(a) and (b), respectively.

As shown in Fig.~\ref{fig:experiment-platform}(a), a three-phase standard current source generates independently adjustable phase currents, which pass through the magnetic sensor array via a three-core cable. A multi-channel acquisition card simultaneously samples reference current signals from the high-accuracy current sensors integrated in the standard current source, and output voltage signals of all magnetic field sensors. The sampled data is transmitted to a LabVIEW-based host computer for subsequent processing and analysis.

Fig.~\ref{fig:experiment-platform}(b) displays the physical test platform, where the magnetic sensor array comprises eight tunneling magnetoresistance (TMR) sensors with a product No. TMR2102~\cite{tmr2102} equally spaced on a 35\,mm-radius circumference. The operational principle of each magnetic field sensor is illustrated in the inset at the bottom-left of Fig.~\ref{fig:experiment-platform}(b). The differential output signals from the TMR sensing elements are first converted to single-ended signals via an instrumentation amplifier AD8220, then processed through a low-pass filter to obtain the output voltage $V_0$, which is subsequently sampled by the multi-channel data acquisition card.

During current measurements, the test current frequency is set to 10\,Hz to eliminate interference from ambient power-frequency (50/60\,Hz) magnetic fields, enabling a more precise evaluation of the proposed self-error correction method. This strategy of employing a specific frequency to avoid power-frequency interference is commonly adopted in laboratory instrument calibration. In practical applications, if interference from external magnetic fields of the same frequency is severe, appropriate magnetic shielding methods must be employed to guarantee measurement accuracy, such as the design of a shielding structure\cite{Zhu2017onsite}. The 16-bit multi-channel acquisition card features a ±5\,V input voltage range and its sampling rate is set 100\,Sa/s. Phase-locking is performed on non-integer period data (e.g., 0.101\,s) to extract fundamental voltage phasors from each channel. This methodology enables the acquisition of dynamically varying phase data within the $[-\pi,\pi]$ range, thereby establishing a robust foundation for data-driven modeling of phase dynamics. In contrast, integer-period phase-locking would yield essentially static phase information, where observed variations stem solely from acquisition card sampling jitter or voltage signal noise.

When all magnetic field sensors operate under normal error conditions, the transfer matrix $\mathbf{K}$ in (\ref{eq:current-inversion}) is first calibrated. To determine the coefficients between phase-a current $\dot{I}_1$ and individual magnetic field sensor $\dot{V}_n$, the three-phase standard current source outputs an 1\,A sinusoidal current through phase a. The reference current signal of phase a and the output voltage signals of all magnetic field sensors are synchronously acquired. Through phase-locking processing, the fundamental current phasor and each magnetic field sensor's fundamental voltage phasor are extracted. The corresponding coefficients are then calculated as follows
\begin{equation}
 k_{n1} = \frac{\dot{V}_{n}}{\dot{I}_{1}}
\end{equation}

The coefficients for the remaining two phase currents with respect to each magnetic field sensor can be determined using a similar methodology, thereby completing the calibration of the magnetic-array-type current sensor.

According to (\ref{eq:current-inversion}), the uncertainty components primarily arise from the transfer coefficient matrix and the uncertainty of the fundamental voltage phasor of the magnetic field sensors. The elements in the transfer coefficient matrix are determined by the relative position between the magnetic field sensors and the measured conductor, as well as the sensitivity of the magnetic field sensors. A "fix-then-calibrate" procedure is adopted. After installation, the uncertainty introduced by the relative position between the magnetic field sensors and the conductor is neglected. The temperature coefficient of the magnetic field sensors is $0.06\%/^{\circ}\mathrm{C}$. With the laboratory temperature range of $20^{\circ}\text{C}$ to $30^{\circ}\text{C}$, and assuming a uniform distribution, the relative standard uncertainty contributed by sensitivity is 0.17\%. The built-in current sensor of the standard current source has an accuracy class of 0.05, contributing an uncertainty of $0.05\%/\sqrt{3}\approx 0.029\%$ to the transfer coefficient matrix.

The uncertainty in the output voltage signal of the magnetic field sensors stems from the magnetic field sensor's intrinsic noise, the multi-channel acquisition board, and the Fourier transform. The fundamental amplitude error due to noise does not exceed 10\,mV, following a uniform distribution. When the measured three-phase current has an RMS value of 10\,A, the relative standard uncertainty of the measured current, evaluated using the Monte Carlo method, is less than 0.8\%.

\subsection{Training and Testing Data Preparation}\label{sec:III-B}
\begin{figure}[!t]
\centering
\includegraphics[width=1\columnwidth]{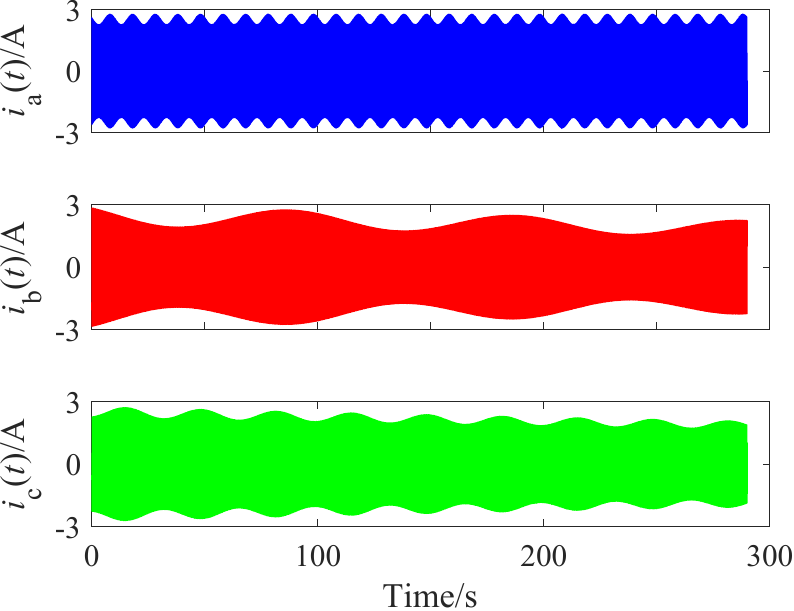} 
\caption{Modulated three-phase currents in time domain for self-error correcting experiment.} 
\label{fig:modulated-current-t}
\end{figure}

\begin{figure}[!t]
\centering
\includegraphics[width=1\columnwidth]{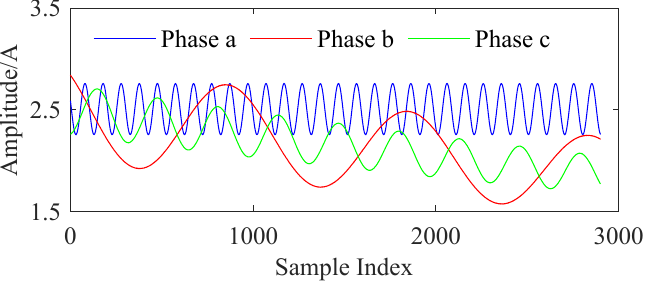} 
\caption{Amplitude variation curves of the modulated three-phase currents for self-error correcting experiment.} 
\label{fig:modulated-current}
\end{figure}

While performing the self-error correcting experiments for the magnetic-array-type current sensor, the three-phase standard current source generates modulated output three-phase currents with an $120^\circ$ phase difference. The modulated three-phase currents in time domain are presented in Fig.~\ref{fig:modulated-current-t}. The current amplitude of phase a follows a sinusoidal modulation, while phases b and c currents exhibit sinusoidal decay variation in current amplitude with distinct exponential attenuation coefficients. It can be seen that the three-phase currents are strictly asymmetric, and the linear correlation between them changes drastically over time. The fundamental current phasors are computed every 0.101\,s and the amplitude variations of three-phase currents are shown in Fig.~\ref{fig:modulated-current} with a total of 2900 sampling points.

\begin{figure*}[!t]
    \centering
\includegraphics[width=1\textwidth, height=0.5\textheight, keepaspectratio]{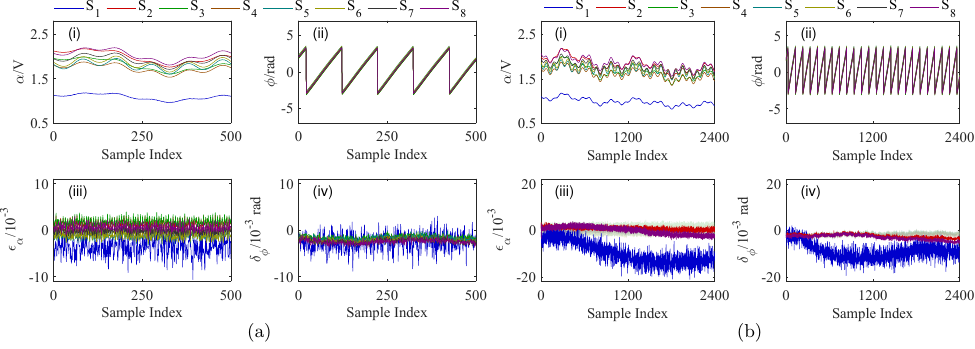} 
\caption{Training data and testing data. (a) Measurements of all magnetic field sensors operating in their normal error status for self-error correcting model training, (b) Measurements of all magnetic field sensors with $\mathrm{S}_1$, $\mathrm{S}_2$, and $\mathrm{S}_8$ having an error drift for self-error correcting method validation. Subplots (i) and (ii) represent the fundamental voltage amplitude and phase, respectively, for each magnetic field sensor. Subplots (iii) and (iv) show the relative error and phase error, respectively, of each magnetic field sensor, derived from the output reference currents of the standard three-phase current source and the calibrated magnetic field-current coefficients matrix using (\ref{eq:1}).}
\label{fig:train-test-data}
\end{figure*}

In the first 500-sample period, all magnetic field sensors are working in a normal error status and their output fundamental voltage phasors served as the training data. During the period of samples 501-2900, the corresponding output voltage phasors of all magnetic field sensors constitute testing data with certain magnetic field sensors exhibiting an error drift.

Using (\ref{eq:1}), the reference voltage signal for each magnetic field sensor can be obtained based on the measured reference currents and the pre-calibrated coefficient matrix $\mathbf{K}$, making it available for monitoring the error drift of each magnetic field sensor. When all magnetic field sensors are in a normal error status, the amplitudes and phases of their fundamental voltage components as training data are shown in Fig.~\ref{fig:train-test-data}(a) (i) and (ii), respectively. It can be observed that the variation trends of the fundamental voltage amplitudes are generally consistent across all magnetic field sensors, and the phases exhibit a sawtooth-like variation within the range of $[-\pi,\pi]$.

The relative error (denoted as $\epsilon_\alpha$) and phase error (denoted as $\delta_\phi$) are obtained to visualize the error drift of each magnetic field sensor, as shown in Fig.~\ref{fig:train-test-data}(a) (iii) and (iv). From Fig.~\ref{fig:train-test-data}(a) (iii), due to its lower sensitivity and poorer signal-to-noise ratio, magnetic field sensor $\mathrm{S}_1$ exhibits relative and phase error in the ranges of $(-10\sim-2)\times10^{-3}$ and $(-8\sim3)\times10^{-3}$\,rad, respectively. In contrast, the relative error and phase error of the other magnetic field sensors are confined within $(-3\sim3)\times10^{-3}$ and $(-3\sim0)\times10^{-3}$\,rad, respectively. The fundamental amplitude and phase data of all magnetic field sensors, as shown in Fig.~\ref{fig:train-test-data} (a) (i) and (ii), are used as training data for data-driven modeling of the amplitude and phase of each magnetic field sensor.

During the testing phase, the fundamental amplitude and phase of the output voltages for all magnetic field sensors are shown in Fig.~\ref{fig:train-test-data}(b) (i) and (ii), respectively. To emulate the error drifts of certain magnetic field sensors during long-term operation, $\text{S}_1$, $\text{S}_2$, and $\text{S}_8$ are heated to alter their sensitivity. The resulting shift trends in relative error and phase error for all magnetic field sensors are illustrated in Fig.~\ref{fig:train-test-data}(b) (iii) and (iv), respectively. As shown in Fig.~\ref{fig:train-test-data}(b) (iii), due to the negative temperature coefficient characteristic of the TMR sensing elements, $\text{S}_1$ exhibits the most significant error drift. Its relative error shifts in a negative direction from $-5\times10^{-3}$ toward $-15\times10^{-3}$, and its phase error changes from $-2\times10^{-3}$\,rad to $-12\times10^{-3}$\,rad, as shown in Fig.~\ref{fig:train-test-data} (b)(iv). In comparison, $\text{S}_2$ and $\text{S}_8$ display smaller error drifts, both trending toward negative values with drift amounts of $-2\times10^{-3}$\,rad and $-4\times10^{-3}$\,rad, respectively. The core objective of following work is to identify the magnetic field sensors presenting error drifts and accurately quantify the error drift amounts, thereby enabling self-error correction of the magnetic-array-type current sensor.

\subsection{Evaluating Measurement Error Status of Magnetic-Array-Type Current Sensor}\label{sec:III-C} 

\begin{figure}[!t]
\centering
\includegraphics[width=0.75\columnwidth, keepaspectratio]{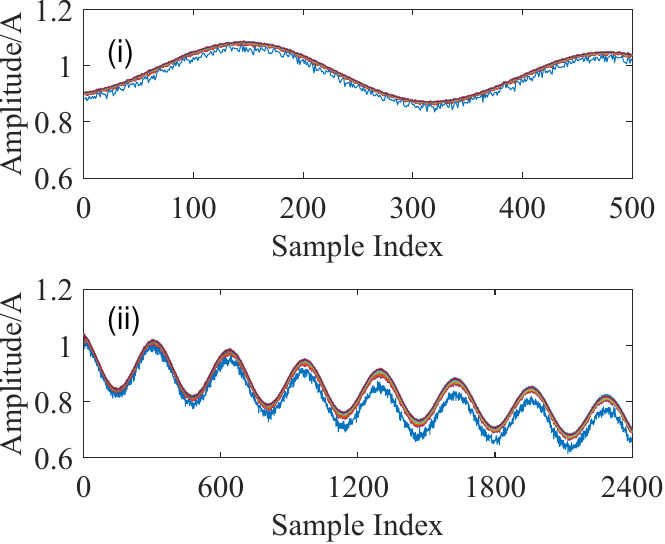} 
\caption{Comparison of linear correlations in phase c current amplitudes for magnetic field sensors under normal and error-drifted conditions. Subplots (i) and (ii) show the amplitudes of the calculated phase c currents derived from the training data and testing data, respectively, of $\mathrm{C}_N^6$
 sub-combinations of the magnetic field sensors.} 
\label{fig:curr-rela}
\end{figure}

Measurement error status of the magnetic-array-type current sensor can be evaluated by detecting the linear correlation variation of multiple fundamental current phasors of per phase, with detailed method illustrated in Section \ref{sec:II-C}. To illustrate the variation in the linear correlation of the calculated currents corresponding to the different sub-combinations of magnetic field sensors caused by error-drifted ones, the sensors are grouped into sub-combinations of six. Using the fundamental voltage amplitude and phase of each sub-combination, as shown in Fig.~\ref{fig:train-test-data} (a)(i) and (ii), the fundamental current phasors per phase are calculated according to (\ref{eq:current-inversion}). Taking phase c as an example, a total number of $\mathrm{C}_{N}^6$ fundamental current phasors are obtained, with their amplitudes presented in Fig.~\ref{fig:curr-rela}(i). Since all magnetic field sensors are working in their normal error status, the linear correlation between the obtained fundamental current amplitudes are essentially consistent over the training period.

Using the fundamental voltage amplitudes and phases of all magnetic field sensors shown in Fig.~\ref{fig:train-test-data} (b)(i) and (ii) as testing data, the magnetic field sensors are again grouped into sub-combinations of six. According to (\ref{eq:current-inversion}), a total number of $\mathrm{C}_N^6$ fundamental current phasors for each phase are obtained. The amplitudes of the obtained fundamental current phasors for phase c are illustrated in Fig. ~\ref{fig:curr-rela}(ii). Compared with Fig.~\ref{fig:curr-rela}(i), due to the error drifts in $\text{S}_1$, $\text{S}_2$, and $\text{S}_8$, deviations arise among the fundamental current amplitudes calculated from different magnetic field sensor sub-combinations, indicating a change in their linear correlation.

\begin{figure*}[!t]
\centering
\includegraphics[width=1\textwidth,height=20cm,keepaspectratio]{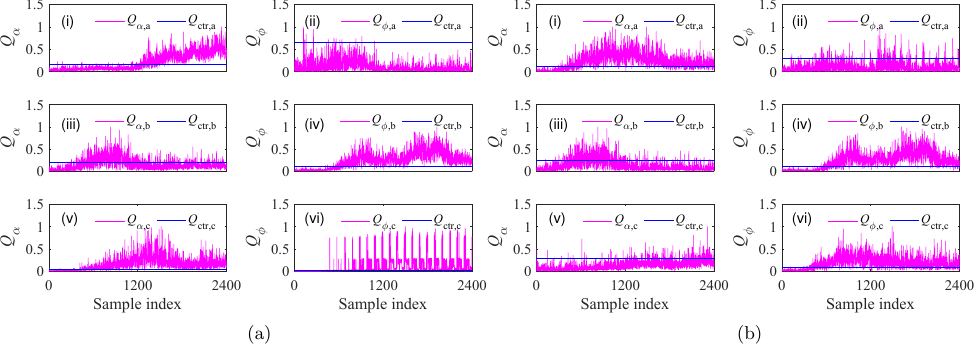} 
\caption{$Q$ statistics and their control thresholds for evaluating the linear correlations of multiple fundamental current phasors per phase. $Q_{\alpha,j}$ and $Q_{\phi,j}$ ($j=\mathrm{a,b,c}$) represents $Q$ statistics for fundamental current amplitude and phase for each phase, respectively and $Q_{\mathrm{ctr},j}$ denotes the corresponding $Q$ statistic threshold. Since it is only necessary to evaluate whether the $Q$ statistic exceeds its control threshold, without needing to know its specific magnitude, the $Q$ statistic associated with its control threshold is normalized to the range of [0,1], the $Q$ statistic curves in the subsequent sections are processed in the same manner. (a) $Q$ statistics for per-phase fundamental current amplitude and phase derived from $C_N^3$ sub-combinations of magnetic field sensors, (b) $Q$ statistics for per-phase fundamental current amplitude and phase derived from $C_{N-1}^3$ sub-combinations of magnetic field sensors with $S_8$ excluded from the magnetic sensor array.} 
\label{fig:error_sensor_identification}
\end{figure*}

To accurately evaluate the measurement error status of the magnetic-array-type current sensor, all magnetic field sensors are grouped into sub-combinations with three sensors each, and a total number of $\mathrm{C}_N^3$ fundamental current phasors for data-driven modeling and testing are obtained. To detect the linear correlation of fundamental current phasors, a data-driven model is constructed for the amplitudes and phases of the $\mathrm{C}_N^3$ fundamental current phasors. According to (\ref{eq:Q-calculation}), the $Q$ statistics for both fundamental current amplitudes and phases calculated from the testing data for each phase are obtained, as shown in Fig.~\ref{fig:error_sensor_identification}(a).

As observed from Fig.~\ref{fig:error_sensor_identification}(a), the $Q$ statistic for the amplitude of phase a remained below the corresponding control threshold during the first half of the test period but exceeds it in the latter half. The $Q$ statistic for the phase of phase a remains within the control limits. This shows that the error drifts of magnetic field sensors might affect only the linear correlation of the amplitudes or the phases of the fundamental current phasors. For phases b and c, their $Q$ statistics for both amplitude and phase all significantly exceed their control thresholds. Through the determination of the $Q$ statistic, it can be sensitively assessed whether magnetic field sensors with an error drift are present in the magnetic-array-type current sensor. For any given phase, if the $Q$ statistic for the amplitudes or phases of the fundamental phase currents derived from different sub-combinations significantly exceeds their respective control thresholds, it can be concluded that error drift has occurred in one or more magnetic field sensors within the magnetic-array-type current sensor.

\subsection{Identifying Error-Drifted Magnetic Field Sensors by Comparing $Q$ Statistic Curves Variation Trend}\label{sec:III-D} 
Upon detecting an abnormal error status in the magnetic-array-type current sensor, all magnetic field sensors presenting an error drift can be further identified by applying the methodology outlined in Section \ref{sec:II-E}, which includes two steps.

First, all $N$ magnetic field sensors are grouped into $C_{N}^3$ sub-combinations, from which the same number of fundamental current phasors for each phase can be obtained based on (\ref{eq:current-inversion}). Fig.~\ref{fig:error_sensor_identification}(a) shows the $Q$ statistic curves for both amplitude and phase of the resulting per-phase fundamental current phasors. These $Q$ statistic curves are used as a reference to identify the error status of each magnetic field sensor by $Q$ statistic curves variation trend comparison.

Second, by iteratively excluding one magnetic field sensor from the magnetic sensor array, the remaining $N-1$ magnetic field sensors are grouped into $C_{N-1}^3$ sub-combinations. From these sub-combinations, $C_{N-1}^3$ fundamental current phasors per phase are obtained. The corresponding $Q$ statistics for both amplitude and phase are then derived using (\ref{eq:Q-calculation}). Fig.~\ref{fig:error_sensor_identification}(b) shows the $Q$ statistics obtained by excluding $\mathrm{S}_8$ from the magnetic sensor array. The $Q$ statistic curves in subplots (i), (v), and (vi) of Fig.~\ref{fig:error_sensor_identification}(b) exhibit significantly different variation trends compared to their corresponding reference curves in Fig.~\ref{fig:error_sensor_identification}(a). The variation trends in subplots (iii) and (iv) of Figs.~\ref{fig:error_sensor_identification}(a) and (b) are fundamentally similar. From the above $Q$ statistic variation trends comparison, the error drift in $\mathrm{S}_8$ significantly alters the linear correlation of the phase-a current amplitude, as well as the amplitude and phase of the phase-c currents. If other magnetic field sensors operating normally are individually excluded, the resulting $Q$ statistic variation trends are similar to those in Fig.~\ref{fig:error_sensor_identification}(a) and these results are not given for clarity.

The measurement error status of each magnetic field sensor can be individually identified following the above procedure. This process provides a foundation for quantifying the error drift amounts in Section~\ref{sec:III-E}.

\subsection{Quantifying Error drifts of Magnetic Field Sensors by Solving a Bi-objective Optimization Problem}\label{sec:III-E} 

Upon identifying the error status of all magnetic field sensors, their error drift magnitudes are evaluated according to the methodology presented in Section \ref{sec:II-E}. As a representative case, we evaluate the error drift amount of magnetic field sensor $\text{S}_1$ using the testing data from Fig.~\ref{fig:train-test-data}(b). The fundamental voltage amplitude $\alpha_t$ and phase $\phi_t$ for $\text{S}_1$ are compensated by their corresponding error drift amounts ($\Delta\alpha_t$ and $\Delta\phi_t$, respectively), i.e., 
\begin{equation}
   \left\{
   \begin{array}{l}
   \alpha_{t}^{\prime} = \alpha_{t} + \Delta\alpha_{t} \\
   \phi_{t}^{\prime} = \phi_{t} + \Delta\phi_{t}
   \end{array}
   \right.
\end{equation}
where $\alpha_{t}^{\prime}$ and $\phi_{t}^{\prime}$ are respectively compensated fundamental voltage amplitude and phase.

A new magnetic sensor array is constructed by combining $\text{S}_1$ with all the magnetic field sensors working in a normal error status and is then organized into multiple sub-combinations, with each consisting of three magnetic field sensors. The fundamental current phasors for each sub-combination can be derived using (\ref{eq:current-inversion}) . Specifically, for sub-combinations containing $\text{S}_1$, the fundamental current phasor computation utilizes the compensated output fundamental voltage parameters ($\alpha_{t}^{\prime}$ for amplitude and $\phi_{t}^{\prime}$ for phase) of the error-drifted magnetic field sensor.

The $Q$ statistics for the amplitude $Q_{\alpha,j}$ and phase $Q_{\phi,j}$ of all fundamental current phasors per phase are computed via (\ref{eq:Q-calculation}) and aggregated to form the composite metrics $Q_\alpha$ and $Q_\phi$ represented by (\ref{eq:bi-objective}).

When the compensated error drift of a magnetic field sensor precisely equals its actual value, the linear correlation among fundamental current phasors derived from the compensated magnetic field measurements remains essentially unaffected, resulting in the minimization of the $Q$ statistics $Q_\alpha$ and $Q_\phi$. Thus, quantification of the error drift amount of $\mathrm{S}_1$ is naturally converted to optimize the defined bi-objective optimization problem in (\ref{eq:bi-objective-problem}). In the case study of evaluating amplitude and phase error drift of $\mathrm{S}_1$ at its 1000th sampling instant, the solution of this bi-objective optimization yields the Pareto front and the corresponding Pareto set are respectively shown in  Fig.~\ref{fig:Pareto_results}(i) and (ii). Specially, the values of both objective functions in Fig.~\ref{fig:Pareto_results}(i) are normalized to a range of [0,1]. During solving the bi-objective optimization problem using NSGA, the population size and iteration number are respectively set as 300 and 50. Solving a single data point takes 4 seconds on a computer with a 12th Gen Intel(R) Core(TM) i7-12700F (2.10 GHz) processor and 16 GB of memory. In practical measurements, with readings taken every 15 minutes, this is sufficient for self-error correction. However, if faster response times are required, further research is necessary into more efficient algorithms for solving the bi-objective optimization problem, such that the proposed method can be integrated into a microcomputer system for industrial or smart grid applications.

\begin{figure}[!t]
\centering
\includegraphics[width=1\columnwidth, keepaspectratio]{figure11.pdf} 
\caption{Bi-objective optimization results for self-error correction of magnetic field sensor $\mathrm{S_8}$ at its 1000th sampling point. The $Q$ statistics for both amplitude $\mathrm{Q}_\alpha$ and phase $\mathrm{Q}_\phi$, are normalized to the range of [0,1]. (a) Pareto front, (b) Pareto set. The magenta pentagram in (a) indicates the closest point to the coordinates, formed by the bi-objective optimal values, and its corresponding Pareto-optimal solution, denoted as a magenta pentagram in (b), is chosen as the final solution.} 
\label{fig:Pareto_results}
\end{figure}

As shown in Fig.~\ref{fig:Pareto_results}(ii), the final solution is graphically denoted by a pentagram marker. This final solution's bi-objective coordinates, marked by the corresponding pentagram in Fig.~\ref{fig:Pareto_results}(i), demonstrate minimal Euclidean separation from the utopia point (square marker) formed by the individual objective minima.

\begin{figure*}[h]
\centering
\includegraphics[width=1\textwidth, height=0.5\textheight, keepaspectratio]{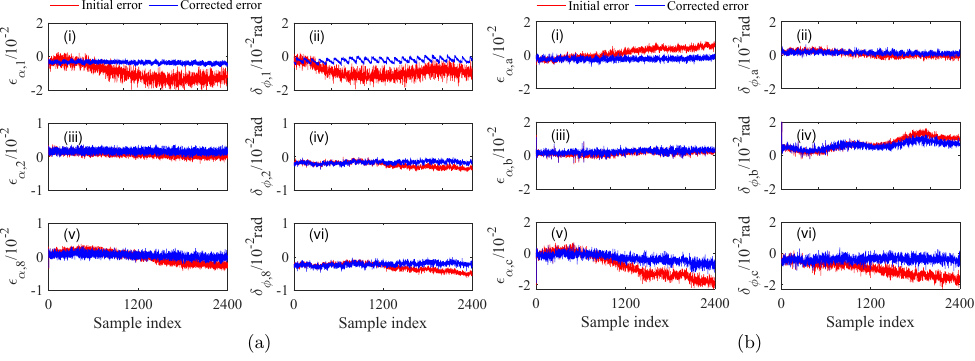} 
\caption{Initial error and corrected error for both magnetic field sensors and derived per-phase currents using the proposed method in the paper. (a) The initial and corrected values for relative error $\epsilon_{\alpha,s}$ ($s=1,2,8$) and phase error $\delta_{\phi,s}$ of magnetic field sensors $\mathrm{S}_1$, $\mathrm{S}_2$, and $\mathrm{S}_8$, (b) The initial and corrected values for relative error $\epsilon_{\alpha,j}$ ($j=\mathrm{a,b,c}$) and phase error $\delta_{\phi,j}$ of fundamental current amplitude and phase per phase.} 
\label{fig:corrected_err}
\end{figure*}

\begin{figure}[!t]
\centering
\includegraphics[width=0.75\columnwidth, keepaspectratio]{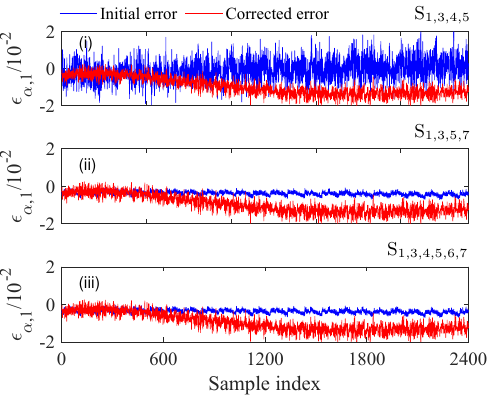} 
\caption{Self-error correction results for relative error of $\rm{S}_1$ using different combinations. (i) Corrected results for combination ($\rm{S}_1$, $\rm{S}_3$, $\rm{S}_4$, $\rm{S}_5$), (ii) Corrected results for combination ($\rm{S}_1$, $\rm{S}_3$, $\rm{S}_5$, $\rm{S}_7$), and (iii) Corrected results for combination ($\rm{S}_1$, $\rm{S}_3$, $\rm{S}_4$, $\rm{S}_5$, $\rm{S}_6$, $\rm{S}_7$)} 
\label{fig:corrected_err_S1comb_}
\end{figure}

\begin{figure*}[h]
\centering
\includegraphics[width=1\textwidth, height=0.5\textheight, keepaspectratio]{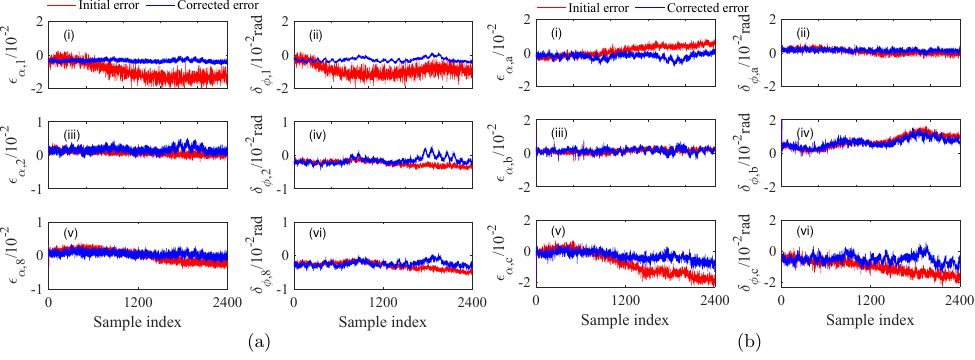} 
\caption{Initial error and corrected error for both magnetic field sensors and derived per-phase currents using the PCA-based multi-latent variable method. (a) The initial and corrected values for relative error $\epsilon_{\alpha,s}$ ($s=1,2,8$) and phase error $\delta_{\phi,s}$ of magnetic field sensors $\mathrm{S}_1$, $\mathrm{S}_2$, and $\mathrm{S}_8$, (b) The initial and corrected values for relative error $\epsilon_{\alpha,j}$ ($j=\mathrm{a,b,c}$) and phase error $\delta_{\phi,j}$ of fundamental current amplitude and phase per phase.} 
\label{fig:corrected_err_noSolve}
\end{figure*}

The amplitude and phase error drifts of magnetic field sensors $\mathrm{S}_1$, $\mathrm{S}_2$, and $\mathrm{S}_8$ are individually determined using the aforementioned methodology.  Fig.~\ref{fig:corrected_err}(a) presents the comparative results of relative error and phase error before and after compensation. As evidenced by Fig.~\ref{fig:corrected_err}(a), the compensated relative and phase errors of $\mathrm{S}_1$, $\mathrm{S}_2$, and $\mathrm{S}_8$ demonstrate significant restoration to their nominal error-state levels. It can be seen from Fig.~\ref{fig:corrected_err}(a) that both the relative error and phase error of $\rm{S}_8$ can be accurately detected and corrected even when their drifts reach values as low as $-2\times10^{-3}$ and $-2\times10^{-3}$\,rad, respectively. This demonstrates a high error drift detection sensitivity of the proposed method for magnetic field sensors.

It should be noted that the number of magnetic field sensors working in a normal status and their placement will have an impact on the error drift quantification performance. Experiments demonstrate the magnetic field sensors in their normal status are preferably distributed uniformly around the multi-core cable to get higher error drift quantification accuracy. Fig.~\ref{fig:corrected_err_S1comb_} (i), (ii) and (iii) respectively show the self-error correction results for relative error of $\rm{S}_1$ using combinations ($\rm{S}_3$, $\rm{S}_4$, $\rm{S}_5$), ($\rm{S}_3$, $\rm{S}_5$, $\rm{S}_7$) and ($\rm{S}_3$, $\rm{S}_4$, $\rm{S}_5$, $\rm{S}_6$, $\rm{S}_7$). As shown in Fig.~\ref{fig:corrected_err_S1comb_}(i), when employing the adjacent magnetic field sensor combination ($\rm{S}_3$, $\rm{S}_4$, $\rm{S}_5$) for self-error correction, the corrected results exhibit significant uncertainty. In Fig.~\ref{fig:corrected_err_S1comb_} (ii) and (iii), when using the uniformly distributed magnetic field sensors ($\rm{S}_3$, $\rm{S}_5$, $\rm{S}_7$) across the cable surface, the error corrected results is consistent with those obtained using all magnetic field sensors in a normal error status.

\subsection{Self-Error Correcting of Magnetic-Array-Type Current Sensors}

Upon accurately quantifying the error drift amounts of all magnetic field sensors working in an abnormal error status, the self-error correction of the magnetic-array-type current sensor can be achieved by calculating phase currents using (\ref{eq:current-inversion}) with the compensated measurements of the magnetic sensor array. Fig.~\ref{fig:corrected_err}(b) compares the relative errors ($\epsilon_\alpha$) and phase errors ($\delta_\phi$) of the three-phase currents before and after implementing the self-error correction procedure.

As shown in Fig.~\ref{fig:corrected_err}(b), before self-error correcting, the relative error of phase-a current has a significant positive drift, altering from $-2\times10^{-3}$ to $6\times10^{-3}$. Both the relative error and phase error of phase c show a significant negative drift, finally reaching approximately $-20\times10^{-3}$ and $-20\times10^{-3}$\,rad, respectively. After performing self-error correction, the corresponding relative error and phase error curves are restored to a horizontal state. The phase error of phase a and the relative error of phase b remain essentially unchanged before and after error compensation. However, the variation trend of the phase error in phase b does not revert to a horizontal line after compensation, which may be caused by fluctuations in the inherent noise of the magnetic field sensors.

These results demonstrate that when specific magnetic field sensors undergo an error drift, the relative and phase error drifts exhibit different degrees of deviation across different phases. For instance, the phase error of phase a [Fig.~\ref{fig:corrected_err}(b) (ii)] and the relative error of phase b [Fig.~\ref{fig:corrected_err}(b) (iii)] exhibit slight drifts, with no significant change before and after error correction.

To further validate the efficacy of the proposed method, a PCA-based multi-latent-variable model ~\cite{Liu2024self} is employed for a comparative study. By minimizing the $Q$ statistics for amplitude and phase of the fundamental voltage phasors of the magnetic sensor array, the relative error and phase error before and after correction for $\rm{S}_1$, $\rm{S}_2$, and $\rm{S}_3$ are presented in Fig.~\ref{fig:corrected_err_noSolve}(a). Compared with Fig.~\ref{fig:corrected_err} (a), it is evident that the corrected relative error and phase error exhibit significant fluctuations due to changes in the correlations among the multiple latent variables. Based on the corrected magnetic field measurements, self-error correction is performed for current measurement error. The pre- and post-correction results are depicted in Fig.~\ref{fig:corrected_err_noSolve} (b). Compared with Fig.~\ref{fig:corrected_err} (b) (i) and (vi), it also presents substantial fluctuations in the corrected relative error for phase a and phase error for phase c. This fluctuation arises because the constructed PCA-based multi-latent-variable model fails to track changes in the correlations between multiple latent variables.

\section{Conclusion}
\label{sec:04}
The magnetic-array-type current sensor facilitates synchronous current measurement in individual conductors within multi-conductor systems, such as multi-core cables, providing a reliable and straightforward non-contact sensing solution without requiring modification or adjustment of the measured circuit. To ensure long-term measurement stability, continuous monitoring and data-driven correction of error drift in each magnetic field sensor are essential for this current sensor architecture.

This paper develops a data-driven single-latent-variable model for measurement error monitoring. By decoupling phase currents, the correlation analysis of multi-latent variables (i.e., multi-conductor currents) is transformed into single-latent-variable modeling (individual phase currents). This approach enables the model to maintain stable error evaluation performance even under significant variations in multi-conductor current correlations, such as fluctuations in three-phase current imbalance.

An experimental validation platform is established in this study, in which a certain number of magnetic field sensors are heated to simulate error drift phenomena. The proposed self-error correcting method accurately identifies magnetic field sensors exhibiting error drift. On this basis, a bi-objective optimization-based approach is introduced to quantify the error drift, followed by error compensation to restore each magnetic field sensor’s error state to its normal condition. Furthermore, the current calculation is performed using the compensated magnetic measurements, successfully correcting the overall measurement error of the magnetic-array-type current sensor to a normal level.

The core principle of the self-error correction method is to use the trained model as a reference. Should the magnetic-array-type current sensor have drifted during the training phase, to ensure long-term measurement accuracy, the current sensor must first be calibrated. Subsequently, the self-error correction is performed to guarantee the sustained accuracy of current measurements. When applying the proposed method for current measurement in three-core cables, the magnetic array must comprise at least five magnetic field sensors, with at least three operating in a normal error status. Better self-error correction performance can be obtained when the error-drift-free magnetic field sensors are distributed relatively uniformly around the multi-core cable.
Due to the mechanical stress, the relative position between the magnetic array and the conductor may shift. This alters the transfer coefficient matrix, which consequently affects the accuracy of error assessment. Therefore, an effort should be made to ensure that the magnetic array remains in a fixed position. Otherwise, the transfer coefficient matrix will require recalibration.


\begin{thebibliography}{10}
\bibitem{Liu2025Research1}
C. Liu, D. Chen and Y. Hou, "Research on early three-core power cable high-impedance fault location method based on the spectrum of propagation functions," IEEE Transactions on Power Delivery, vol. 40, pp. 630-640, 2025.
\bibitem{Liu2023Online2}
J. Liu, C. Lee and P. W. T. Pong, "Online short-circuit fault diagnosis in three-core power distribution cable based on magnetic pattern," IEEE Sensors Journal, vol. 23, pp. 21832-21841, 2023.
\bibitem{Ripka2010Electric3}
P. Ripka, "Electric current sensors: a review," Measurement Science and Technology, vol. 21, p. 112001, 2010.
\bibitem{Sun2024High4}
H. Sun, S. Huang and L. Peng, "High-current sensing technology for transparent power grids: A review," IEEE Open Journal of the Industrial Electronics Society, vol. 5, pp. 326-358, 2024.
\bibitem{Zhu2017onsite}
K. Zhu, W. Han, W. K. Lee, and P. W. T. Pong, "On-site non-invasive current monitoring of multi-core underground power cables with a magnetic-field sensing platform at a substation," IEEE Sensors Journal, vol. 17, pp. 1837-1848, 2017.
\bibitem{Luo2024Research13}
R. Luo, Y. Qin, Y. Zhou, F. Li, and R. Wang, "Research on a current reconstruction method of multi-core cable based on surface magnetic field measurements," Progress In Electromagnetics Research M, vol. 127, pp. 31-39, 2024.
\bibitem{Liu2022Coreless7}
X. Liu, W. He, P. Guo, and Z. Xu, "A coreless current probe for multicore cables," IEEE Sensors Journal, vol. 22, pp. 19282-19292, 2022.
\bibitem{Liu2021Nonintrusive10}
X. Liu, W. He, Y. Zhao, Y. Guo, and Z. Xu, "Nonintrusive current sensing for multicore cables considering inclination with magnetic field measurement," IEEE Transactions on Instrumentation and Measurement, vol. 70, p. 9513314, 2021.
\bibitem{Chen2024Contactless}
K. L. Chen and J. H. Chen, "Contactless current measurement for suspended overhead lines using a magnetic field sensor array," IET Generation, Transmission \& Distribution, vol. 18, pp. 1360-1371, 2024.
\bibitem{Chen2022Intelligent}
K. Chen, "Intelligent Contactless Current Measurement for Overhead Transmission Lines," IEEE Transactions on Smart Grid, vol. 13, pp. 3028-3037, 2022.
\bibitem{Wu2019Overhead}
Y. Wu, G. Zhao, J. Hu, Y. Ouyang, S. X. Wang, J. He, F. Gao, and S. Wang, "Overhead transmission line parameter reconstruction for UAV inspection based on tunneling magnetoresistive sensors and inverse models," IEEE Transactions on Power Delivery, vol. 34, pp. 819-827, 2019.
\bibitem{Ma2025Eliptical}
H. Ma, S. Wu, R. Bi, H. Zhang, J. Tian, D. Zhang, and J. Hu, "Elliptical arrays of tunnelling magnetoresistance sensors for rectangular busbar current measurement," High voltage, vol. 10, pp. 807-819, 2025.
\bibitem{George2023Rectangular}
N. George and P. Ripka, "Rectangular busbar with circular sensing part for wideband current measurement," IEEE Transactions on Instrumentation and Measurement, vol. 72, 2023.
\bibitem{Li2021Wideband}
W. Li, G. Zhang, H. Zhong, and Y. Geng, "A wideband current transducer based on an array of magnetic field sensors for rectangular busbar current measurement," IEEE Transactions on Instrumentation and Measurement, vol. 70, p. 9004511, 2021.
\bibitem{Bellina2002Optimization}
F. Bellina, P. Bettini and F. Trevisan, "Optimization analyses of the magnetic measurements on multistrand SC cables," IEEE transactions on applied superconductivity, vol. 12, pp. 1651-1654, 2002.
\bibitem{Li2016High}
Z. Li, S. Yan, W. Hu, Z. Li, Y. Xu, "High accuracy on-line calibration system for current transformers based on clamp-shape Rogowski coil and improved digital integrator," MAPAN, vol 31, pp. 119-127, 2016.
\bibitem{Chen2025New}
Y. Chen, X. Zhang, D. Li, J. Zhou, "A new class of fault detection and diagnosis methods by fusion of spatially distributed and time-dependent features," Journal of Process Control, 2025,146: 103372.
\bibitem{Yu2025Challenges}
J. Yu and Y. Zhang, "Challenges and opportunities of deep learning-based process fault detection and diagnosis: A review," Neural Computing and Applications, vol. 35, pp. 211-252, 2023.
\bibitem{Tong2024novel}
X. Tong, J. Ma, L. Ma, S. Yan, Q. Tang, Z. Teng, and D. Cheng, "A novel prediction method for smart meter error using multiview convolutional neural network," IEEE Sensors Journal, vol. 24, no. 24, pp. 42009–42017, Dec. 2024.
\bibitem{Zhang2023physics}
Y. Zhang, C. Zhang, C. He, H. Li, Q. Chen, P. Guo, C. Cheng, "A physics-information-enabled self-updating method to monitor steady-state error of capacitor voltage transformers," Measurement, vol. 220, 2023, Art. no. 113295.
\bibitem{Kong2022deep}
I.Kong and Z. Ge, "Deep learning of latent variable models for industrial process monitoring," IEEE Transactions on Industrial Informatics, vol. 18, pp. 6778-6788, 2022.
\bibitem{Gand2024Sensor}
A.Gandhimathinathan, C. G. Ananthakrishnan, R. Lavanya, R. Jehadeesan, and P. R. Reddy, "Sensor anomaly detection in nuclear power plant using deep LSTM denoising autoencoder and isolation forest," IEEE Sensors Letters, vol. 8, pp. 1-4, 2024.
\bibitem{Xia2023Measurement}
T. Xia, C. Liu, M. Lei, S. Xia, D. Li, and D. Ming, "Measurement Error Estimation for Distributed Smart Meters Through a Modified BP Neural Network," Frontiers in Energy Research, vol. 10, 2022.
\bibitem{Zhang2017Monitoring}
Z. Zhang, H. Li, D. Tang, C. Hu, and Y. Jiao, "Monitoring the metering performance of an electronic voltage transformer on-line based on cyber-physics correlation analysis," Measurement Science \& Technology, vol. 28, p. 105015, 2017.
\bibitem{Lu2018Data}
B. Lu, J. Stuber and T. F. Edgar, "Data-driven adaptive multiple model system utilizing growing self-organizing maps," Journal of Process Control, vol. 67, pp. 56-68, 2018.
\bibitem{Zhang2019Decting}
C. Zhang, H. Li, J. Yang, M. Chen, and Y. Jiao, "Detecting measurement error drifts of a capacitor voltage transformer on-line and its field application," Measurement Science \& Technology, vol. 30, p. 105109, 2019.
\bibitem{Liu2024self}
X. Liu, K. Ma, J. Liu, W. Zhao, L. Peng, S. Huang, and S. Li, "A self-healing magnetic-array-type current sensor with data-driven identification of abnormal magnetic measurement units," IEEE Transactions on Instrumentation and Measurement, vol. 73, p. 1008912, 2024.
\bibitem{greenacre2022principal}
C.~Labrín and U. Francisco. ``Principal component analysis,''In R for Political Data Science, Chapman and Hall/CRC, pp. 375-393, 2020.
\bibitem{Jackson1979control}
J. E. Jackson and G. S. Mudholkar, "Control Procedures for residuals associated with principal component analysis," Technometrics, vol. 21, pp. 341-349, 1979.
{\bibitem{tmr2102} TMR2102 Datasheets, MultiDimension Technol. Co., Jiangsu, China.}


\end{thebibliography}
\end{document}